\documentclass{svproc}
\usepackage{url}
\usepackage{xspace}
\usepackage{color}
\usepackage{xcolor}
\usepackage{rotating}
\definecolor{light-gray}{gray}{0.8}

\newcommand{\pp}{\ensuremath{\rm pp}\xspace}
\newcommand{\pPb}{p--Pb\xspace}
\newcommand{\PbPb}{Pb--Pb\xspace}

\newcommand{\GeVc}{\ensuremath{{\rm GeV/}c}\xspace}

\newcommand{\TeV}{\ensuremath{\rm TeV}\xspace}

\newcommand{\sqrts}{\ensuremath{\sqrt{s}}\xspace}

\newcommand{\sqrtsNN}{\ensuremath{\sqrt{s_{\rm NN}}}\xspace}
\newcommand{\pt}{\ensuremath{{\it p}_{\rm T}}\xspace}

\newcommand{\Dzero}{\ensuremath{{\rm D}^{0}}\xspace}

\newcommand{\Dstarplus}{\ensuremath{{\rm D}^{*+}}\xspace}
\newcommand{\Dplus}{\ensuremath{{\rm D}^{+}}\xspace}

\newcommand{\cmnt}[1]{}

\newcommand{\Rppb}{\ensuremath{R_{\rm pPb}}\xspace}

\usepackage{lineno}
\usepackage{amssymb}
\renewcommand{\figurename}{Figure}
\begin{document}
\mainmatter              
\title{Study of open heavy-flavour hadron production in pp and \pPb collisions with ALICE}
\titlerunning{Heavy-flavour hadron production in small system with ALICE.}  
%
\author{Preeti Dhankher\inst{*} on behalf of ALICE collaboration}

\institute{Indian Institute of Technology Bombay,
Mumbai-400076, India\\
\email{preeti.dhankher@cern.ch}}

\maketitle              
\begin{abstract}
Measurement of heavy-flavour production in small systems can be used to test the Quantum ChromoDynamic (QCD) models. In this manuscript, the production cross section of D mesons at midrapidity and open heavy-flavour decay muons (HFM) measured at forward rapidity in pp collisions at \sqrts = 5.02 TeV, and open heavy-flavour decay electrons (HFE) measured at midrapidity in pp collisions at \sqrts = 13 TeV with ALICE detector will be presented. The self-normalized yield of open heavy-flavour decay electrons and muons as a function of multiplicity in pp and \pPb collisions will be presented. Finally, the latest results on nuclear modification factor ($Q_{\rm pPb}$) of D mesons and the $\nu_{2}$ of open heavy-flavour decay electrons and muons in \pPb collisions will be discussed.
\end{abstract}
\section{Introduction}\label{intro}
\vspace*{-0.08in}
Heavy quarks (charm and beauty) are effective probes to test perturbative QCD (pQCD) based calculations in pp collisions and to study cold nuclear matter (CNM) effects such as gluon saturation, shadowing, k$_{\rm T}$ broadening and energy loss in CNM in \pPb collisions. Recent observations in pp and \pPb collisions shown remarkable similarities with \PbPb collisions, which might suggest the presence of collectivity. To further explore the origin of the collective-like effects observed in pp and \pPb collisions, the study of open heavy-flavour production as a function of the charged-particle multiplicity naturally links soft and hard processes that occur in the collision and allows one to study their interplay.

In ALICE \cite{bib:ALICE}, open heavy-flavour hadron are either detected directly via the reconstruction of hadronic decays of D mesons  ( \Dzero, \Dplus, \Dstarplus, and D$^{+}_{s}$) and  $\Lambda^{\rm{+}}_{\rm {c}}$ baryon at midrapidity, or indirectly by a single electron at midrapidity, or muon produced at forward rapidity via semi-leptonic decay channel. In addition, $\Lambda_{\rm {c}}$ and $\Xi_{\rm {c}}$ are also reconstructed via semi-leptonic decays at midrapidity.
\vspace*{-0.1in}
\subsection{\bf{Heavy-flavour production in \pp collisions}}
\vspace*{-0.05in}
The non-strange D-mesons (D$^{0}$, D$^{+}$ and D$^{*+}$) production cross sections measured down to low \pt ($\sim$ 0 \GeVc for \Dzero) in pp collision at \sqrts = 5.02 TeV {\cite{bib:Dmeson} is shown in Fig. \ref{fig:DmesonwithModel}. The measurement is compared with different pQCD models using the Fixed Order plus Next-to-Leading Logarithms approach (FONLL) \cite{bib:FONLL}, k$_{\rm T}$-factorization \cite{bib:kT} and General- Mass Variable-Flavour-Number Scheme (GM-VFNS) \cite{bib:GMVFNS}. All the model describe well the experimental data within their large uncertainties. The data is more precise than the model prediction. Therefore, providing strong constraints on their parameters.
\begin{figure}[!h]
\vspace*{-0.15in}
       \begin{minipage}{0.32\hsize} 
       \begin{center}
       \includegraphics[width=0.8\linewidth]{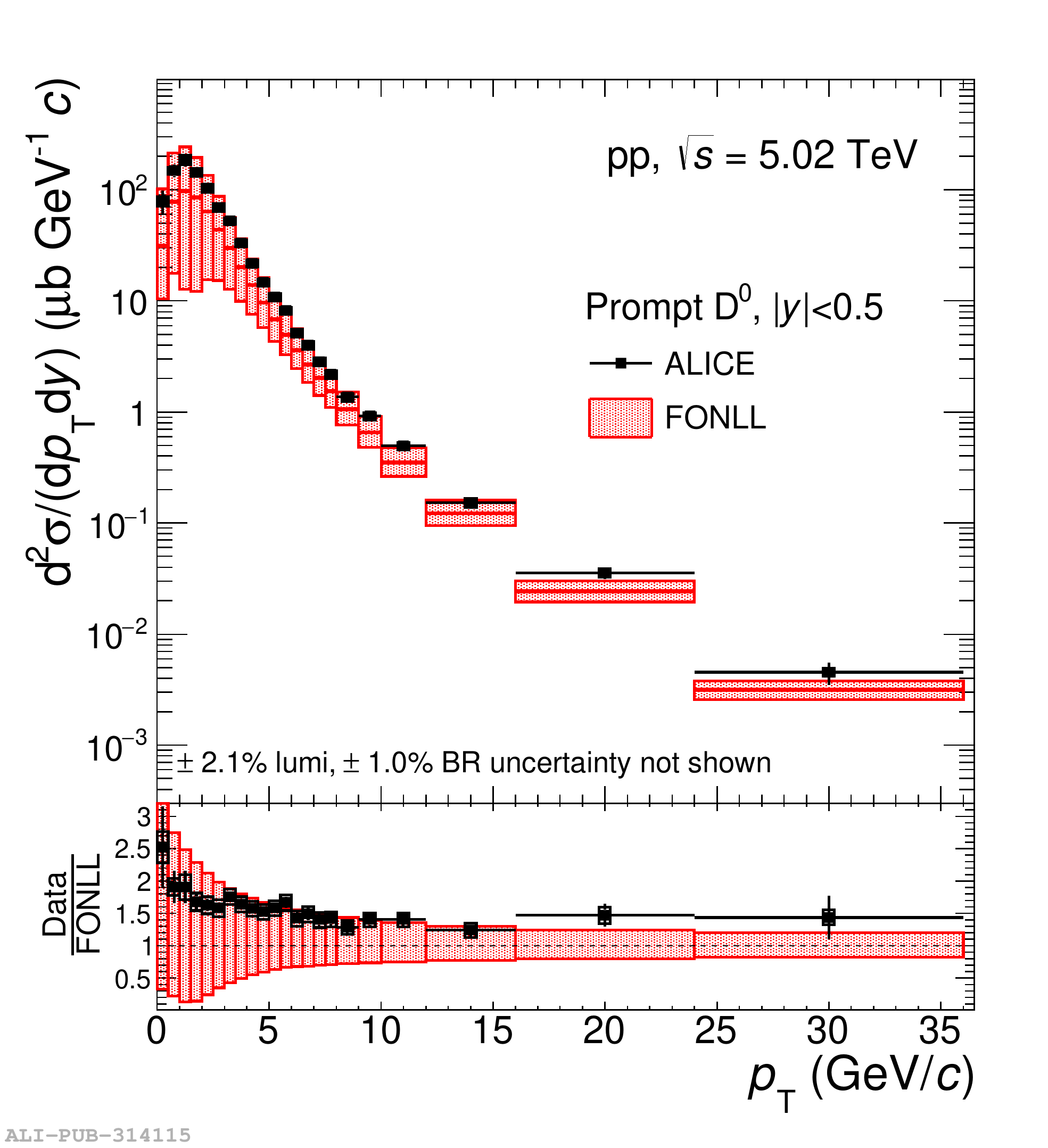}
       \end{center}
       \end{minipage}
        \begin{minipage}{0.32\hsize} 
       \begin{center}
       \includegraphics[width=0.8\linewidth]{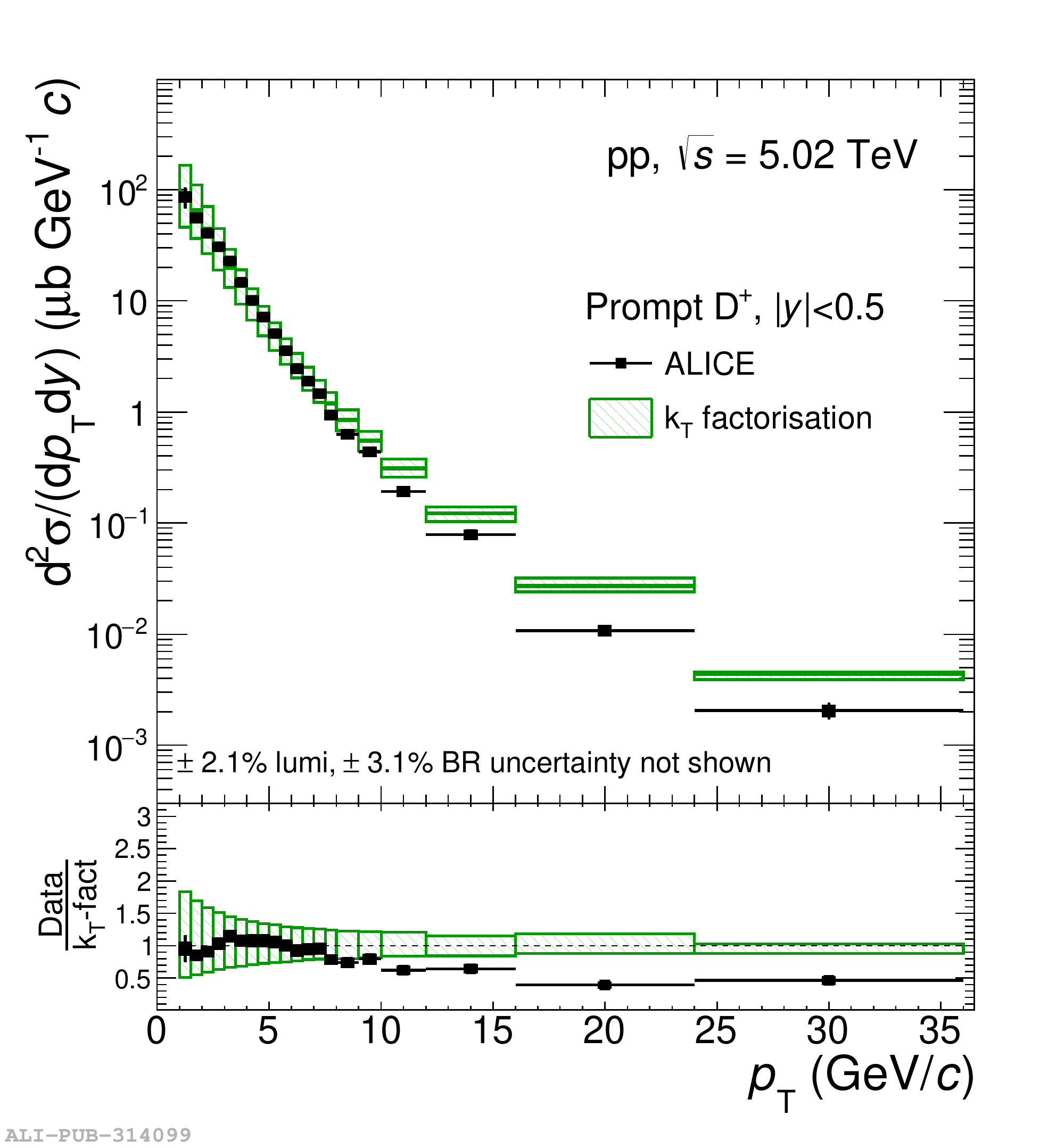}
       \end{center}
       \end{minipage}
       \begin{minipage}{0.32\hsize} 
       \begin{center}
       \includegraphics[width=0.8\linewidth]{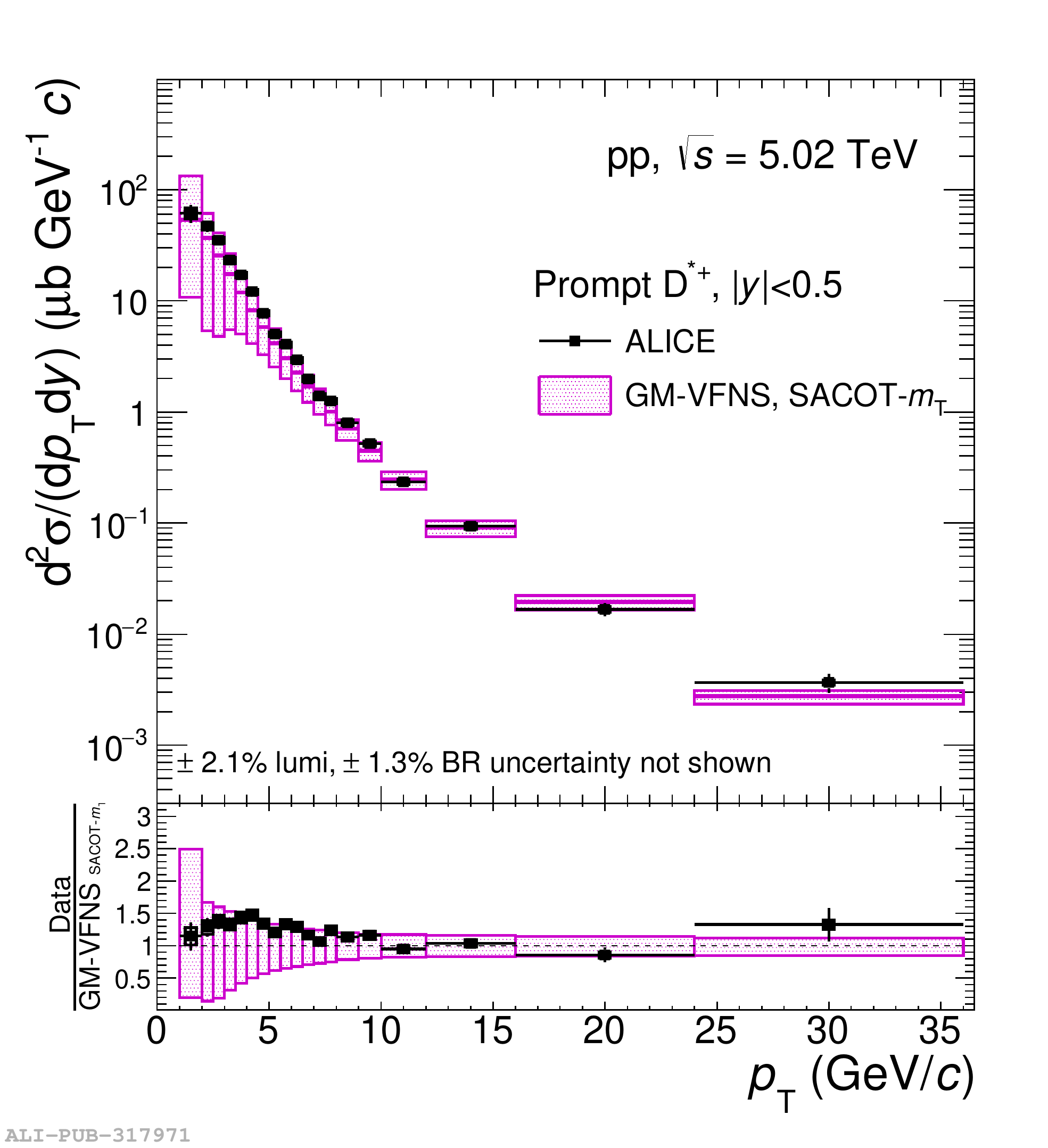}
       \end{center}
       \end{minipage}
         \caption{D-meson production cross section measured in pp collision at \sqrts = 5.02 TeV compared with pQCD models. } 
          \label{fig:DmesonwithModel}
          \vspace*{-0.15in}
       \end{figure}
The production cross section of HFE and HFM ($c,b \rightarrow e/\mu$) in pp collision at \sqrts = 13 TeV and 5.02 TeV is shown in Fig. \ref{fig:SemiLeptonicCrossSection} along with the FONLL model calculations. The data lie at the upper edge of the theoretical prediction for both measurements. The semi-leptonic decay measurement 
for muons (left) also shows separate $c \rightarrow\mu$ and $b \rightarrow \mu$
cross section prediction from FONLL, which indicates that beauty is the main contributor for \pt $\gtrsim$ 5 \GeVc, whereas at low \pt the charm contribution dominates.
\begin{figure}[!h]
 \vspace*{-0.15in}
       \begin{minipage}{0.5\hsize} 
       \begin{center}
       \includegraphics[width=0.58\linewidth]{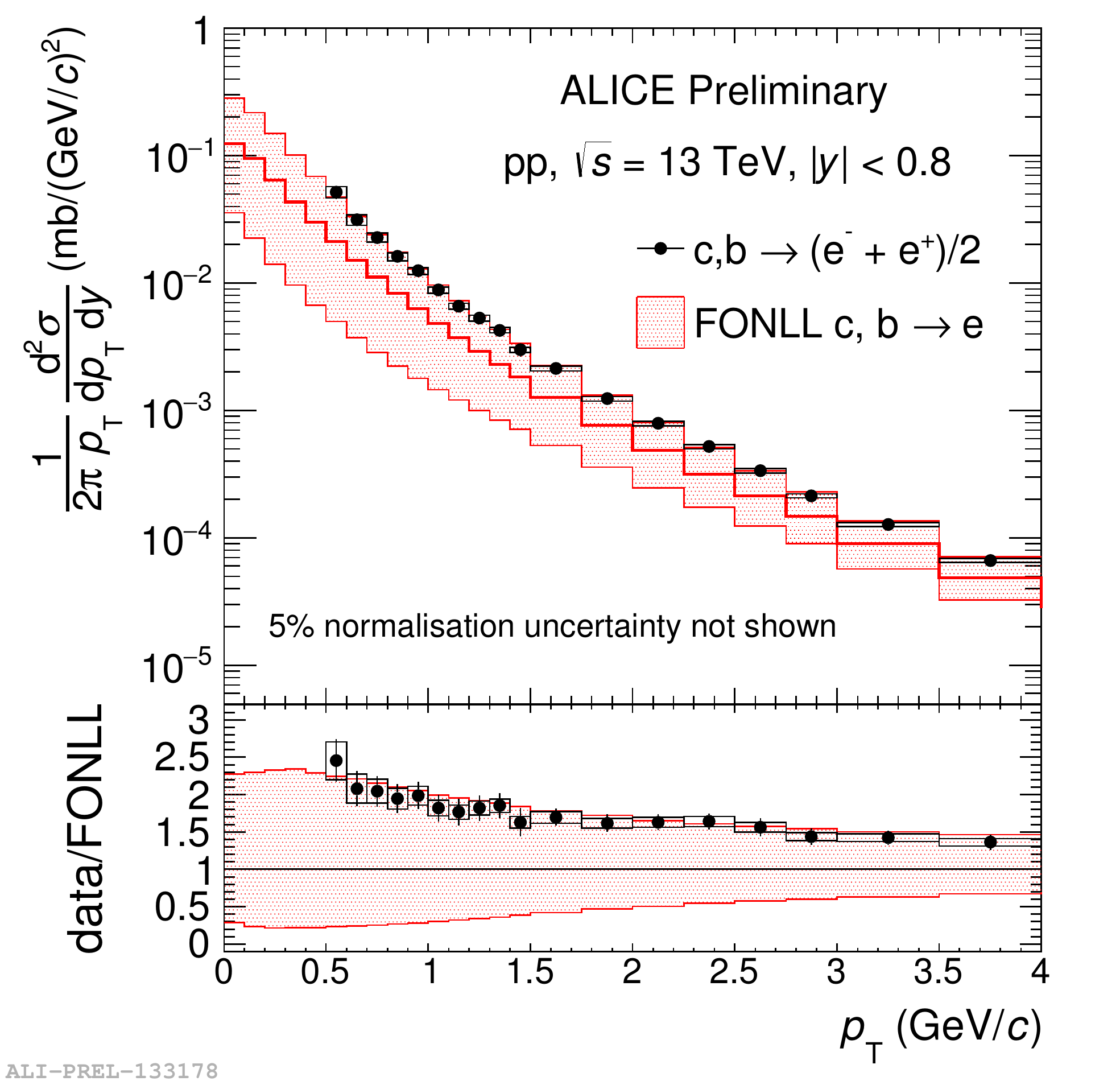}
       \end{center}
       \end{minipage}
       \begin{minipage}{0.5\hsize} 
       \begin{center}
       \includegraphics[width=0.55\linewidth]{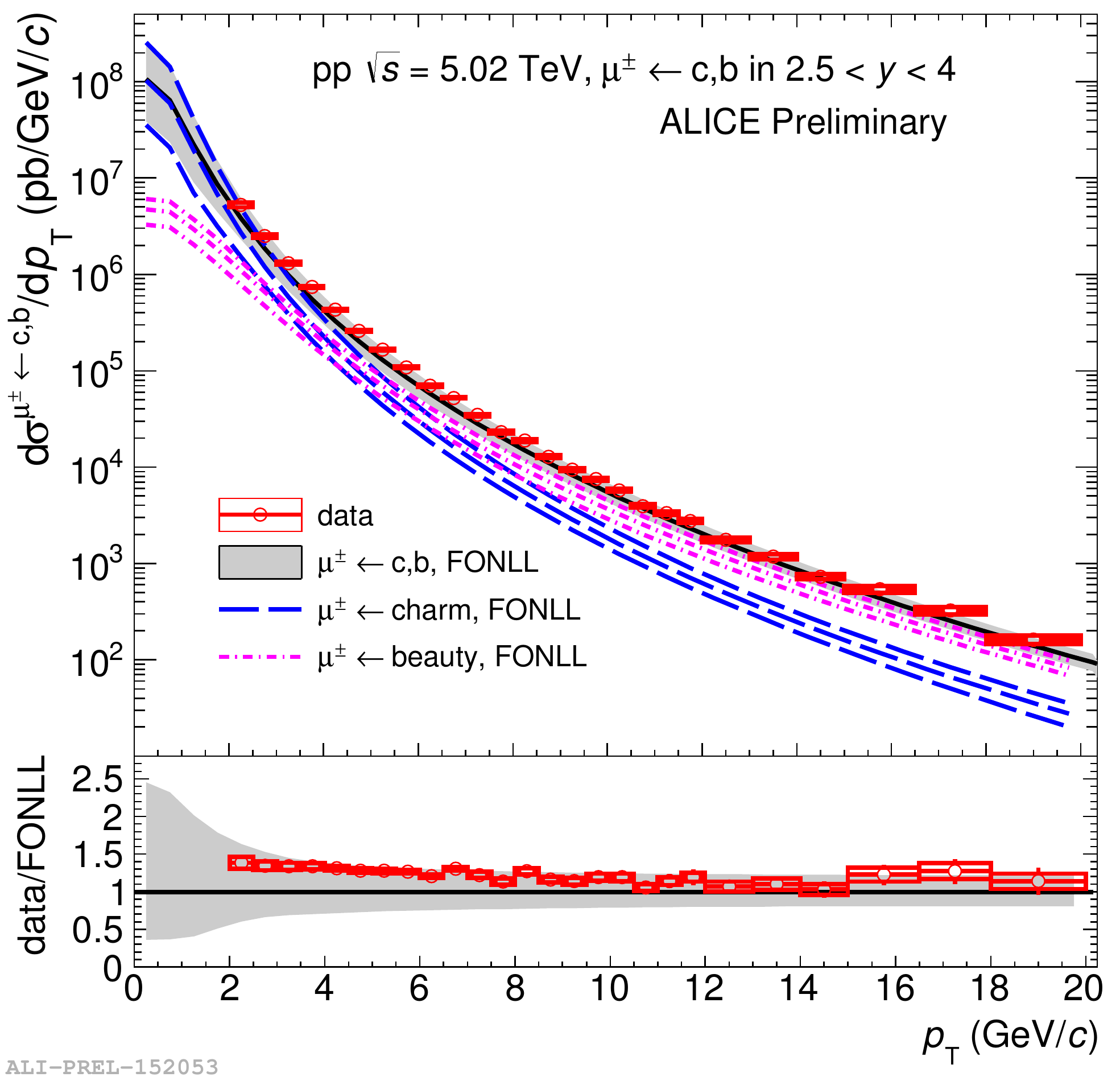}
       \end{center}
       \end{minipage}
         \caption{Production cross section of heavy-flavour hadron decay electrons and muons in pp collision at \sqrts = 13 TeV (left) and 5.02 TeV \cite{bib:Muon} (right). } 
          \label{fig:SemiLeptonicCrossSection}
           \vspace*{-0.15in}
       \end{figure}
        \vspace*{-0.15in}
\subsection{\bf{Multiplicity dependence in heavy-flavour production.}}
The study of heavy-flavour production as a function of multiplicity correlates the hard and soft processes of particle production. The self-normalized yield of HFE and HFM measured in pp collision at \sqrts = 13 TeV (left), and 8 TeV (right) is shown in Fig. \ref{fig:HFSelfNormpp}, in which both exhibit faster-than-linearly increasing trend, where the increase is steeper for high-\pt intervals. The left panel shows the HFE measurement is consistent with the PYTHIA8.2 \cite{bib:PYTHIA} predictions, which incorporates Multiple Parton Interaction (MPI) and indicates that the increasing trend may be linked to MPIs. The righ panel shows HFM measurement compared with the EPOS3.2 \cite{bib:EPOS} predictions that exclude hydrodynamics, which also include MPIs. The data is under predicted by the model at high multiplicity which suggests that there can be some collective-like effects along with the MPI. Figure \ref{fig:HFSelfNormpPb}, left panel, demonstrates that the similar steeper-than-linear trend is observed also for the average D-meson relative yield measured in \pPb collisions and in pp collisions and the right panel shows that the D-meson and HFE measurement are consistent within the uncertainties.
  \begin{figure}[!h]
   \vspace*{-2em}
       \begin{minipage}{0.5\hsize} 
       \begin{center}
       \includegraphics[width=0.8\linewidth]{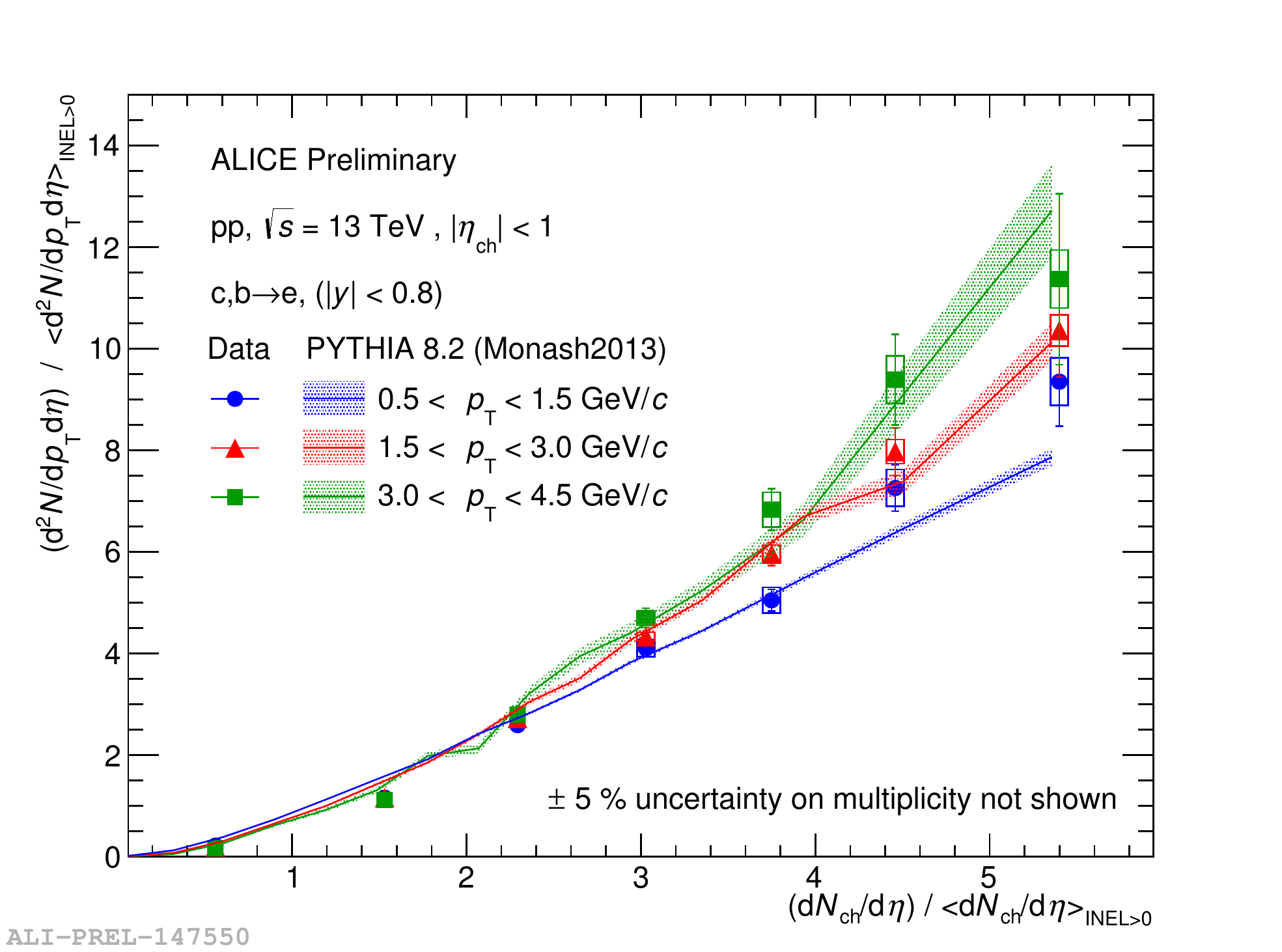}
       \end{center}
       \end{minipage}
       \hspace*{0.01in}
       \begin{minipage}{0.45\hsize} 
       \begin{center}
       \includegraphics[width=0.9\linewidth]{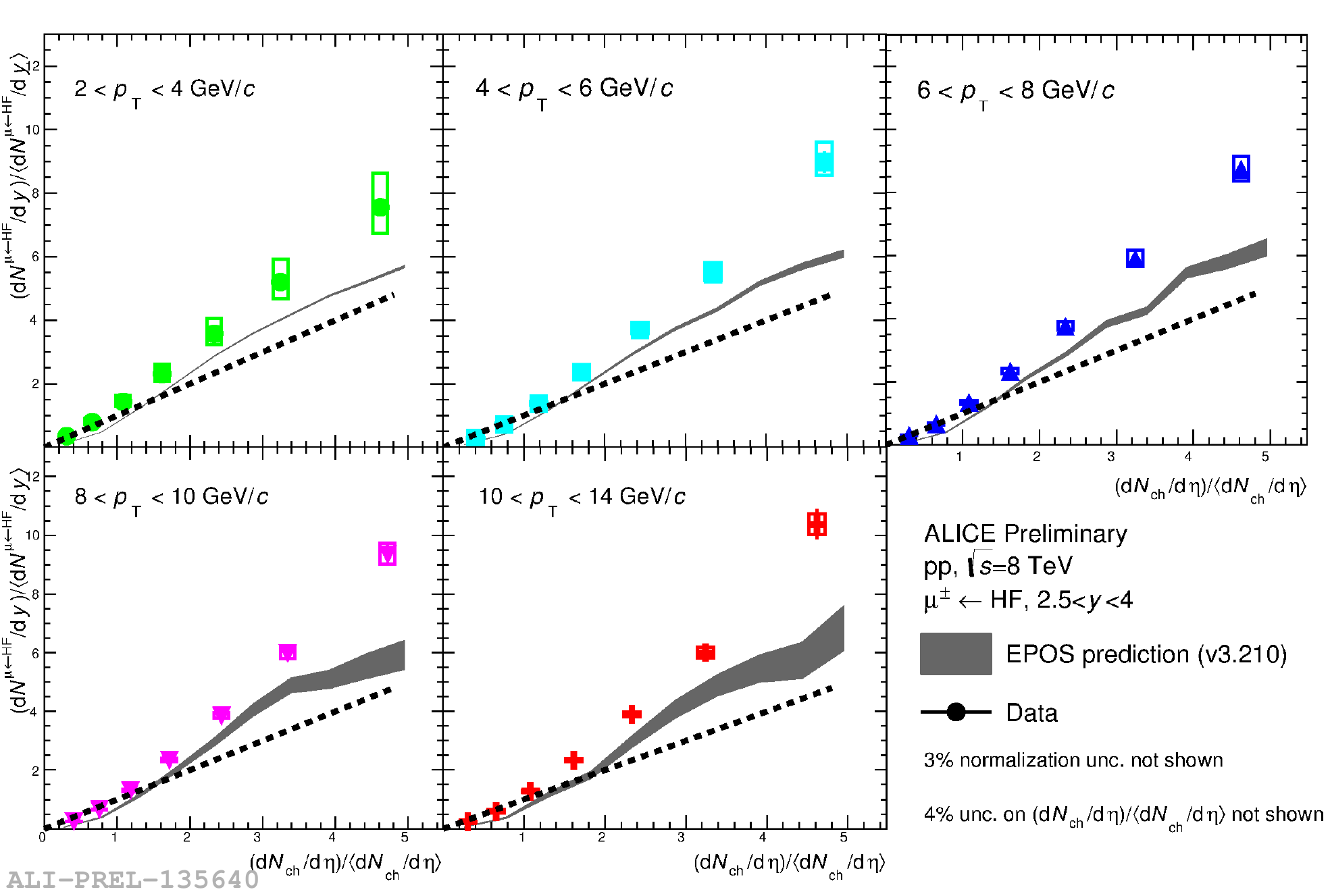}
       \end{center}
       \end{minipage}
         \caption{Multiplicity dependent self-normalised yield of HFE  in \pp collisions at \sqrts = 13 \TeV compared with PYTHIA8.2 predictions (left) and HFM in \pp collisions at \sqrts = 5.02 \TeV compared with EPOS3 without hydrodynamics predictions. } 
          \label{fig:HFSelfNormpp}
          \vspace*{-0.5in} 
       \end{figure}
\begin{figure}[!h]
       \begin{minipage}{0.5\hsize} 
       \begin{center}
       \includegraphics[width=0.6\linewidth]{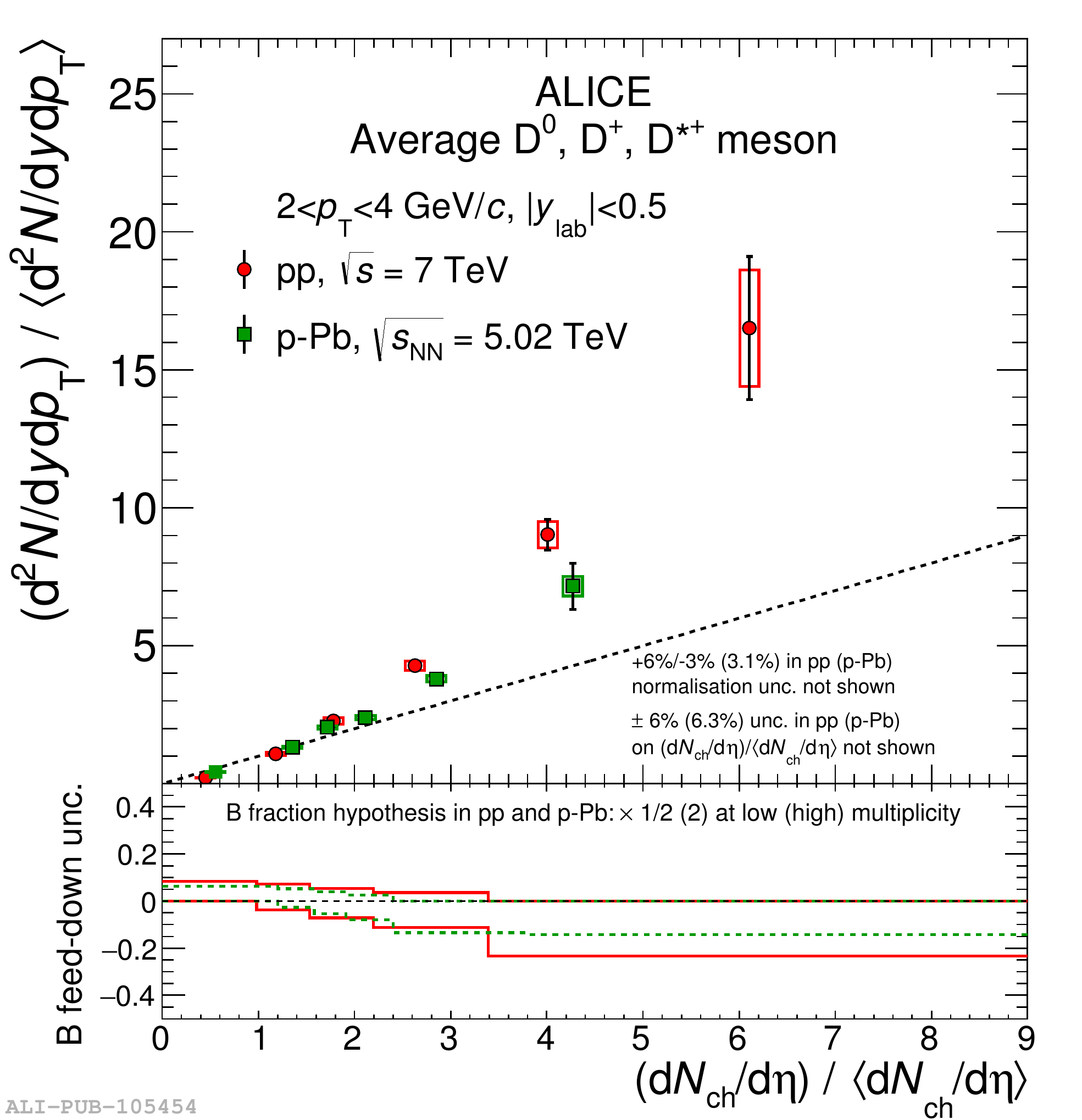}
       \end{center}
       \end{minipage}
       \hspace*{0.01in}
       \begin{minipage}{0.45\hsize} 
       \begin{center}
       \includegraphics[width=0.8\linewidth]{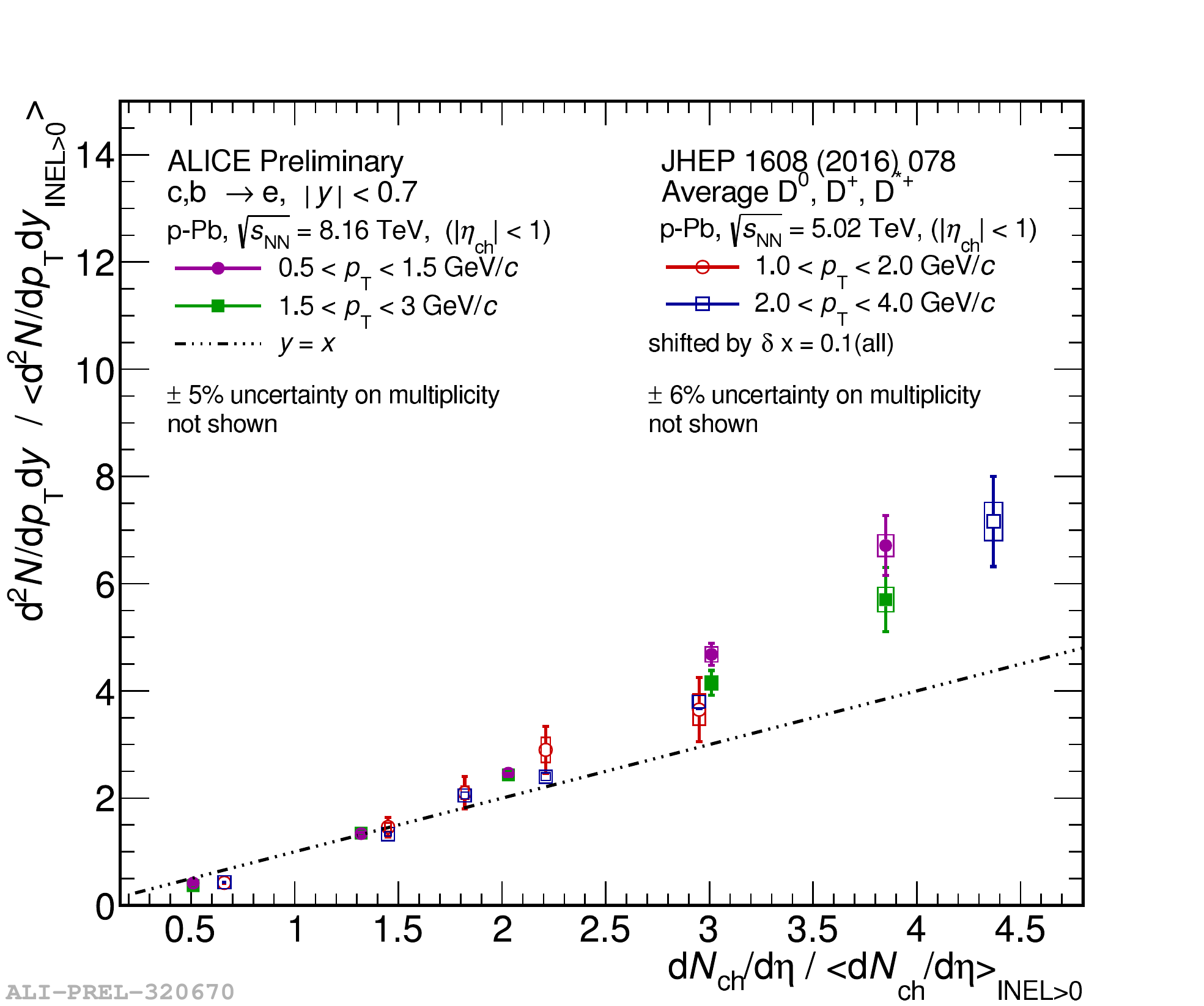}
       \end{center}
       \end{minipage}
         \caption{Multiplicity dependent self-normalised yield of average D-meson in \pp collisions at \sqrts = 7 \TeV(left) and HFE in \pPb collisions at \sqrtsNN = 8.16 \TeV (right) where both are compared with average D-meson measurement in \pPb collisions at \sqrtsNN = 5.02 \TeV. } 
          \label{fig:HFSelfNormpPb}
       \end{figure}
        \vspace*{-0.4in}
\subsection{\bf{Nuclear matter effects and $\it{v}_2$ in \pPb collisions}}
Nuclear modification factor measurements in \pPb collision provide access to Cold Nuclear Matter (CNM) effects and the possible collective-like effects. Figure \ref{fig:DmesonRppb}, left and the middle panel shows the nuclear modification factor (\Rppb) of prompt D-meson given by Eq. \ref{eq:Rppb} and compared with model predictions as well.
\begin{equation} \label{eq:Rppb}
    \Rppb = \frac{1}{A}\frac{{\rm  d}^{2} \sigma_{\rm pPb}^{\rm prompt\ D}/{\rm  d}\pt dy}{{\rm  d}^{2} \sigma_{\rm pp}^{\rm prompt\ D}/{\rm  d}\pt {\rm  d}y},\ Q_{\rm cp} = \frac{({\rm  d}^{2} \sigma_{\rm pPb}^{\rm prompt\ D}/{\rm  d}\pt dy)_{\rm pPb}^{i}}{({\rm  d}^{2} \sigma_{\rm pPb}^{\rm prompt\ D}/{\rm  d}\pt dy)_{\rm pPb}^{60-100\%}}
\end{equation}
The data is described well by the models including the CNM effects \cite{bib:CNM1} \cite{bib:CNM2} \cite{bib:CNM3} \cite{bib:CNM4} left, whereas the middle panel shows the comparison of data with the model which predicts a small QGP formation {\cite{bib:DMesonpPbQGP}} and a suppression at high \pt, which is disfavoured by the data. The right panel shows the measurement of $Q_{\rm cp}$ given by Eq. \ref{eq:Rppb} which is comparable to the measurement of inclusive charged particle. The enhancement observed ( $\sim 3\sigma$) at intermediate \pt could be interpreted as larger radial flow in most central \pPb collisions.
 \begin{figure}[!h]
 \vspace*{-0.1in}
       \begin{minipage}{0.29\hsize} 
       \begin{center}
       \includegraphics[width=0.9\linewidth]{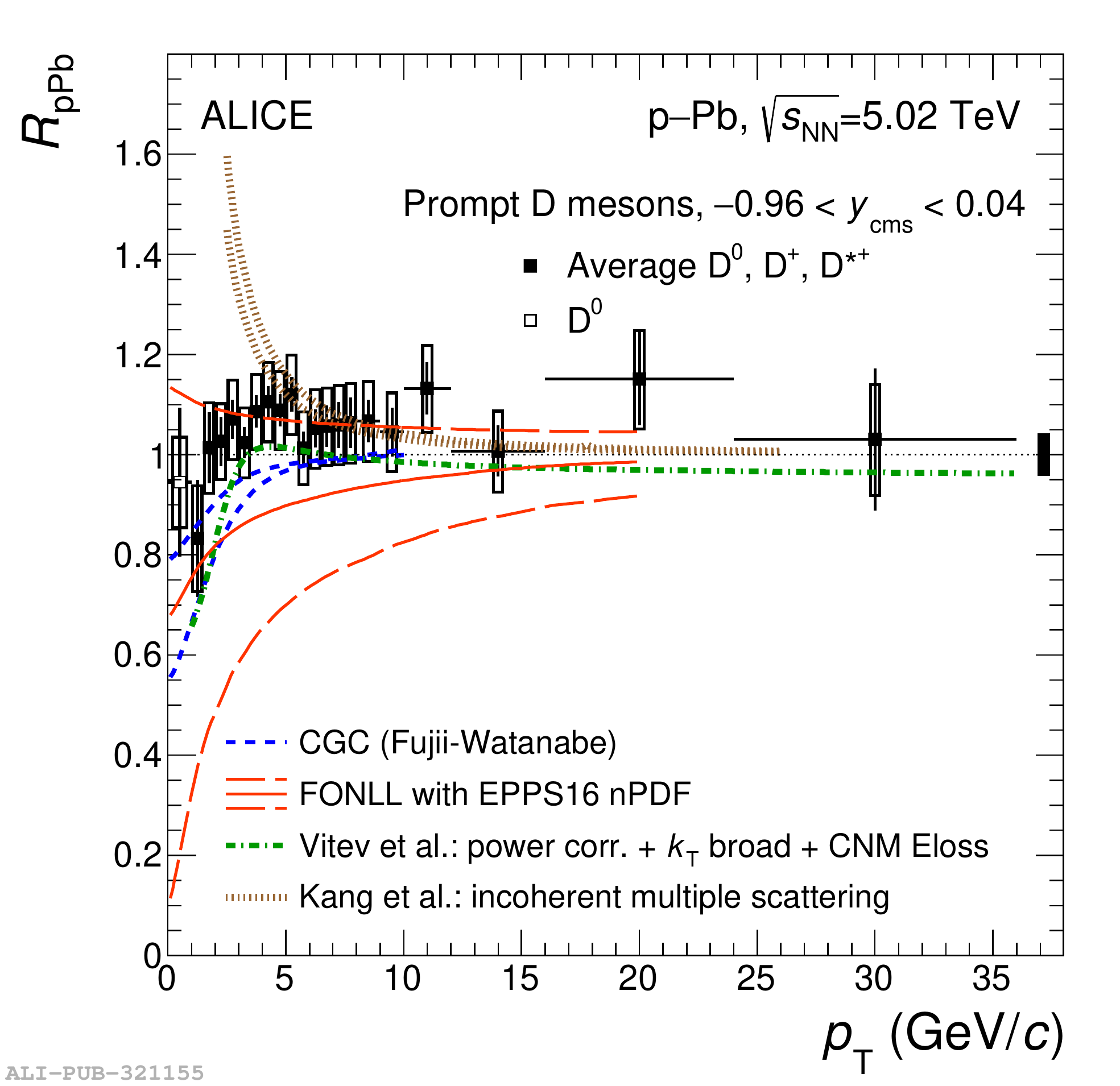}
       \end{center}
       \end{minipage}
        \begin{minipage}{0.29\hsize} 
       \begin{center}
         \includegraphics[width=0.9\linewidth]{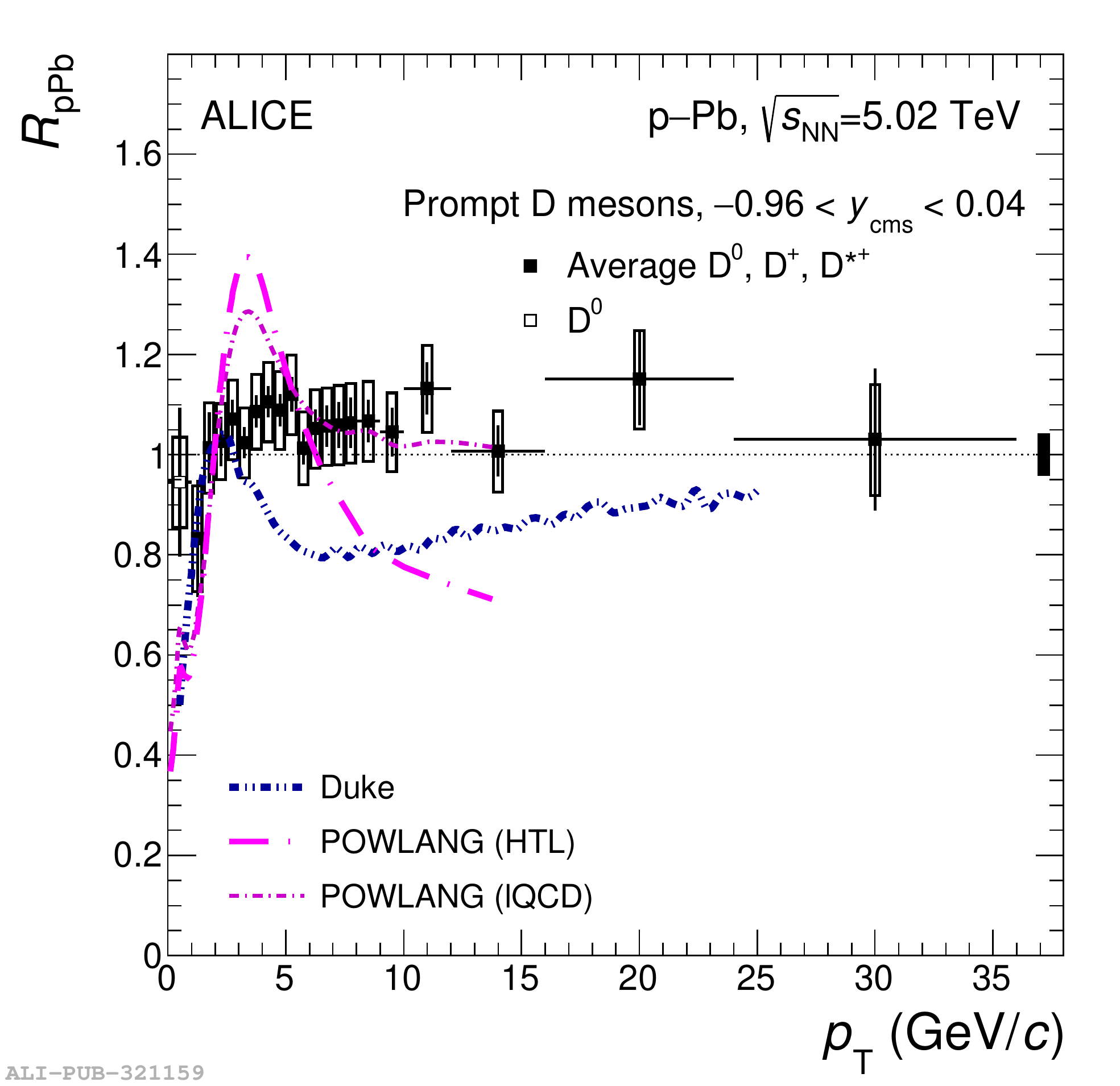}
       \end{center}
       \end{minipage}
       \begin{minipage}{0.41\hsize} 
       \begin{center}
       \includegraphics[width=0.9\linewidth]{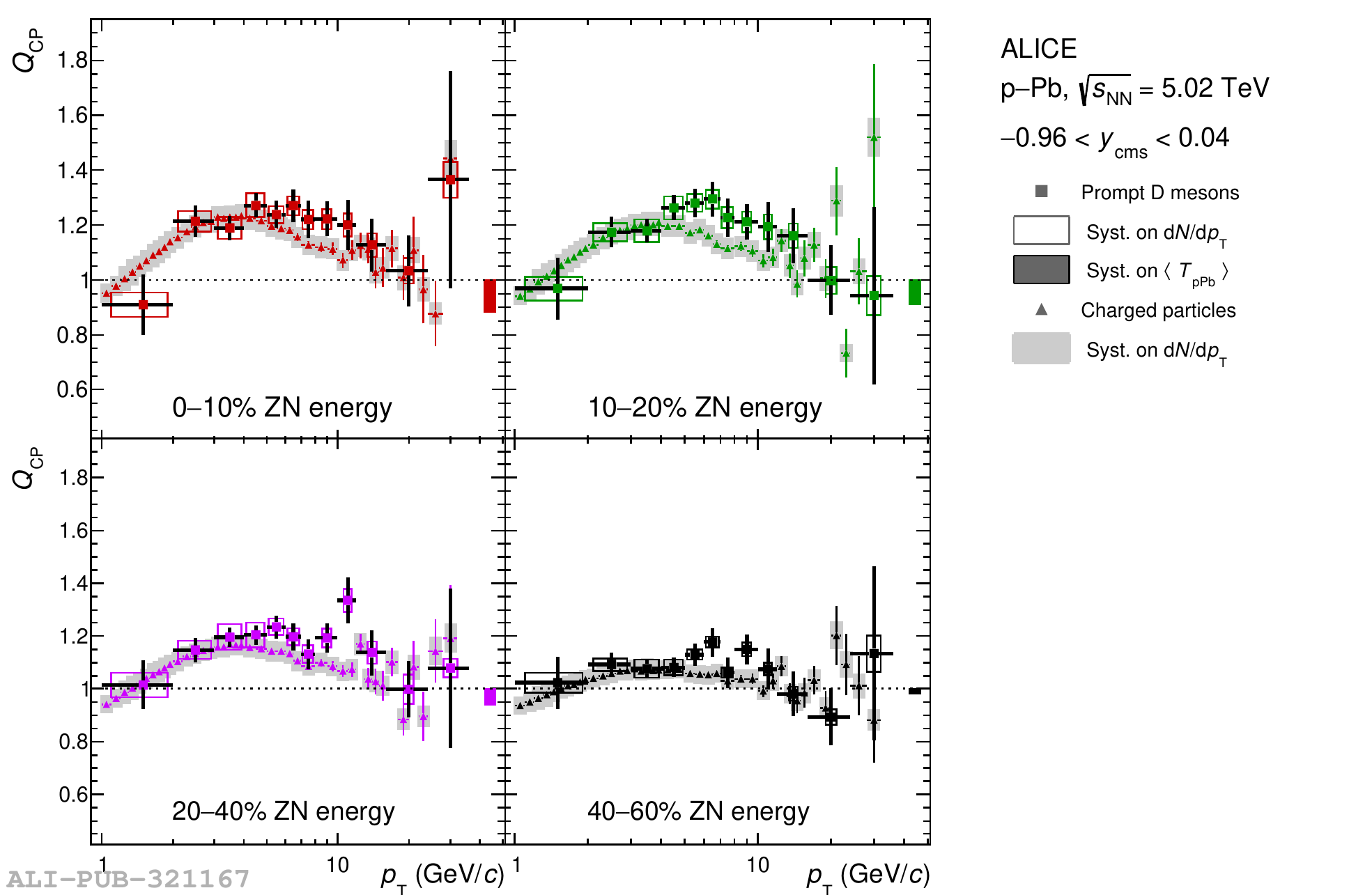}
       \end{center}
       \end{minipage}
         \caption{Nuclear Modification Factor \Rppb (left and middle) and $Q_{\rm cp}$ (right) of prompt D-meson measured in \pPb collisions at \sqrtsNN = 5.02 TeV \cite{bib:DMesonpPb}. \Rppb measurement compared with models including CNM effects (left) and assuming formation of QGP (right).} 
          \label{fig:DmesonRppb}
          \vspace*{-0.2in}
       \end{figure}
As shown in Fig. \ref{fig:EllipticFlow}, a positive elliptic flow coefficient ($\it{v}_{2}$) is observed in the heavy flavour sector, which potentially can be interpreted as the existence of collective like behavior in the small systems. The left panel of Fig. \ref{fig:EllipticFlow} shows $\it{v}_{2}$ measurement, obtained from the decomposition of correlation distribution, of HFE at intermediate \pt with a significance of $\sim 5\sigma$ at \sqrtsNN = 5.02 \TeV \cite{bib:HFEv2} whereas right panel shows $\it{v}_{2}$ measurement of inclusive muons obtained with Q-cumulant method at \sqrtsNN = 8.16 \TeV.
 \begin{figure}[!h]
  \vspace*{-0.2in}
       \begin{minipage}{0.5\hsize} 
       \begin{center}
       \includegraphics[width=0.7\linewidth]{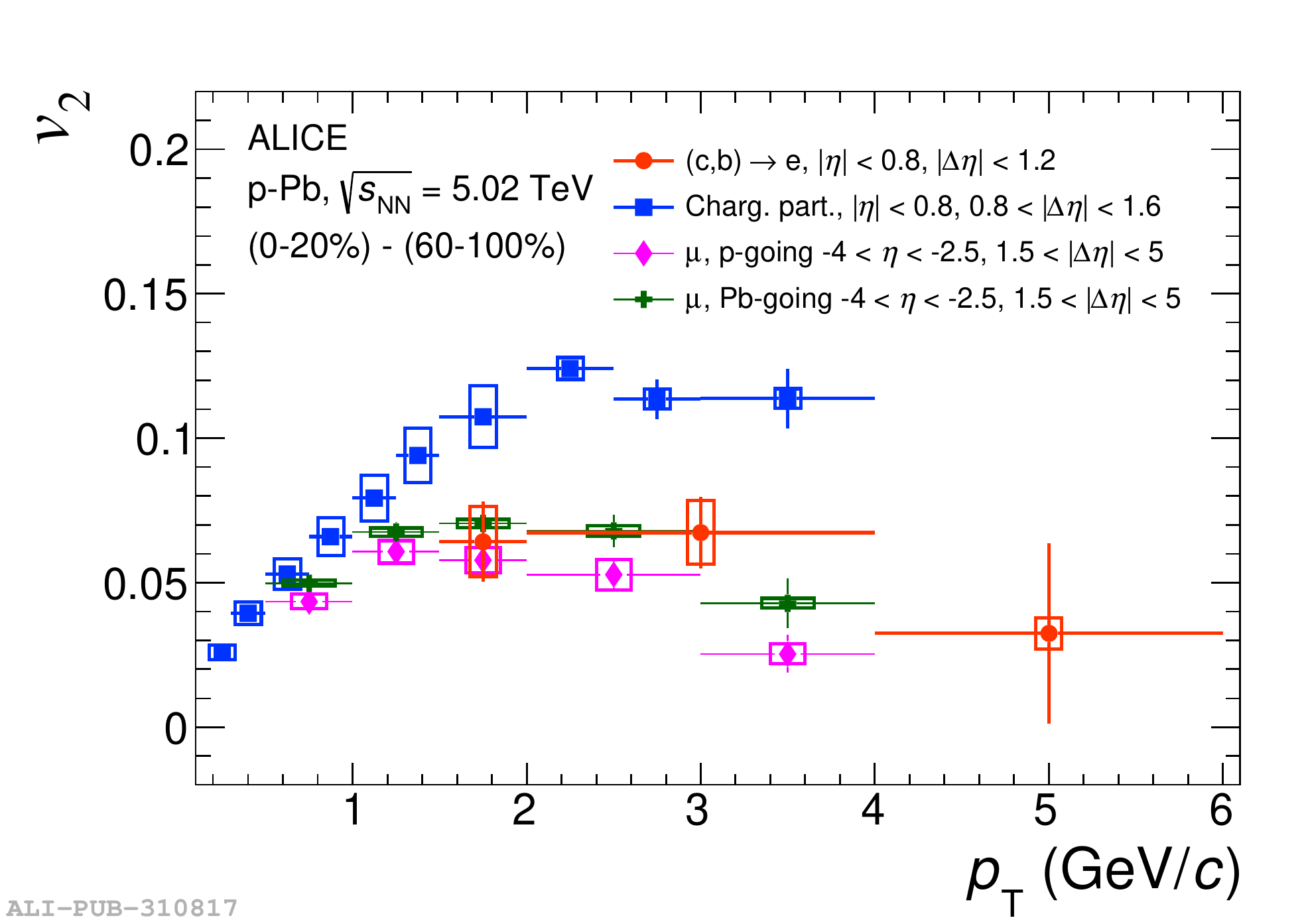}
       \end{center}
       \end{minipage}
       \begin{minipage}{0.45\hsize} 
       \begin{center}
       \includegraphics[width=0.7\linewidth]{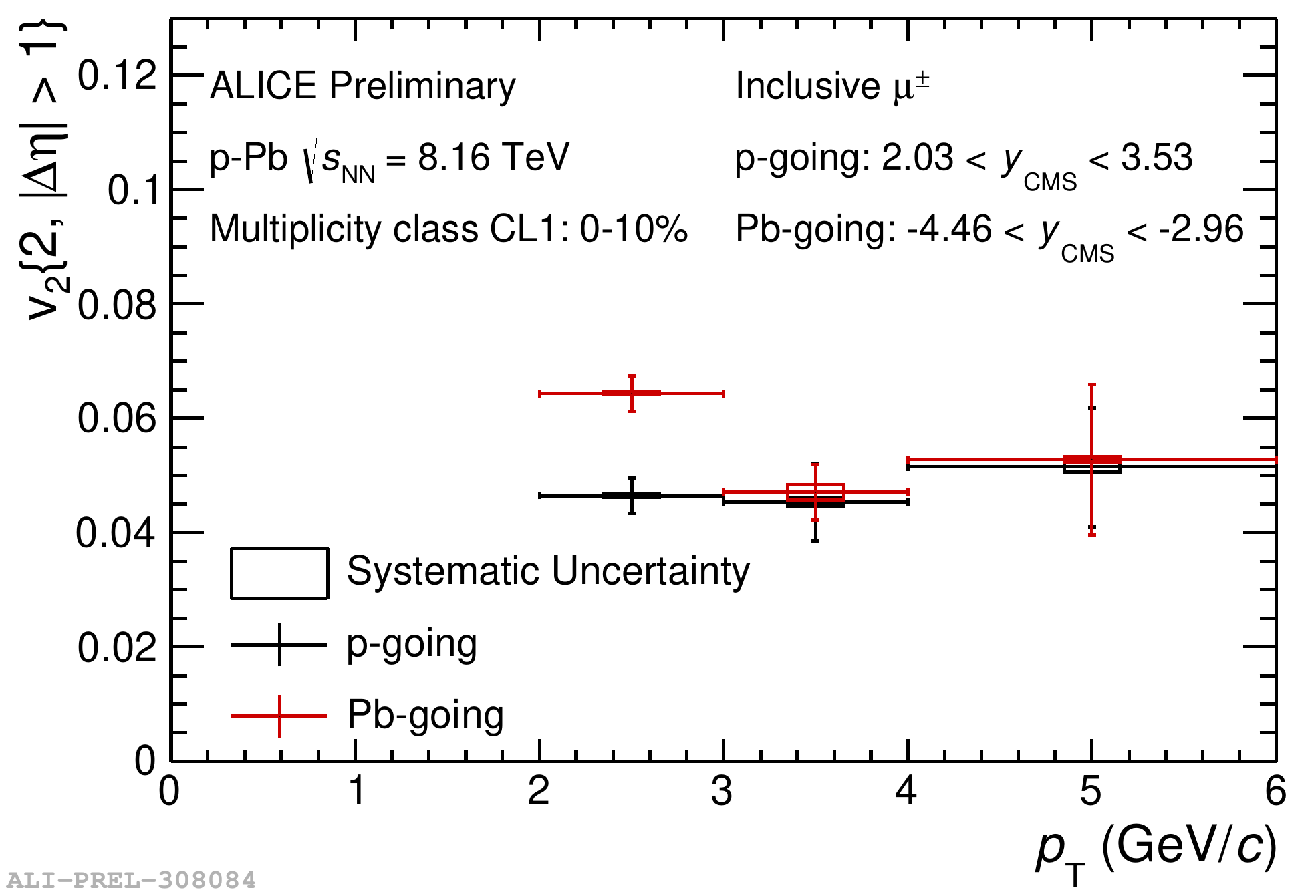}
       \end{center}
       \end{minipage}
         \caption{Elliptic flow coefficient ($\it{v}_{2}$) of HFE compared with charged particle and inclusive muons  measured in \pPb collision at \sqrtsNN = 5.02 TeV \cite{bib:HFEv2} (left) and $\it{v}_{2}$ of inclusive muons measured in \pPb collision at \sqrtsNN = 8.16 TeV (right). } 
          \label{fig:EllipticFlow}
       \end{figure}
   \vspace*{-0.35in}      
\section{Summary and Conclusion}
 \vspace*{-0.1in}
 In summary, the results of heavy-flavour D-meson, HFE, and HFM production cross section measured in pp collision at 5.02 TeV and 13 TeV are discussed. Various pQCD theoretical models describe the \pp data within their uncertainties. The self-normalized yield of D-mesons, HFE, and HFM demonstrates stronger-than-linear dependence on charged-particle multiplicity and this trend can be qualitatively explained by models including MPI. The \Rppb measurement of prompt D-mesons is consistent with no modification over the whole momentum range within the current uncertainties. The $Q_{\rm cp}$ measurement of prompt D-mesons and positive $\it{v}_{2}$ of HFE and HFM in \pPb collisions suggests potential collectivity in the small systems.

\end{document}